# Grazing Incidence Optics for Wide-field X-ray Survey Imaging: A Comparison of Optimization Techniques

Peter W. A. Roming[1,2], John C. Liechty[3], Jared R. Shoemaker[4], David H. Sohn[5], William B. Roush[6], David N. Burrows[1], Gordon P. Garmire[1]

ABSTRACT

Utilizing a ray-tracing program, we have modeled the angular resolution of a short focal length (~2m), large field-of-view (3.1 square degrees), grazing incidence mirror shell. It has been previously shown in the literature that the application of a polynomial to the surface of grazing incidence mirror shells enhances the global performance of the mirror over the entire field-of-view. The objective of this project was to efficiently locate the optimal polynomial coefficients that would provide a 15 arcsec response over the entire field-of-view. We have investigated various techniques for identifying the optimal coefficients in a large multi-dimensional polynomial space. The techniques investigated include the downhill simplex method, fractional factorial, response surface (including Box-Behnken and central composite) designs, artificial neural networks (such as back-propagation, general regression, and group method of data handling neural networks), and the Metropolis-Coupled Markov-Chain Monte-Carlo (MC-MCMC) method. We find of the methods examined, the MC-MCMC approach performs the best. This project demonstrates that the MC-MCMC technique is a powerful tool for designing irreducible algorithms that optimize arbitrary, bounded functions and that it is an efficient way of probing a multi-dimensional space and uncovering the global minimum in a function that may have multiple minimums.

## 1. INTRODUCTION

Optimizations of multidimensional problems are often difficult and require considerable CPU resources. One such problem is the optimization of wide-field x-ray optics. The need for optimizing these wide-field x-ray optics in order to perform deep x-ray surveys is apparent (Chincarini *et al* 1998). Prior investigations into the optimization of grazing incidence optic have already been accomplished.

Utilizing a parabolic-hyperbolic design and a hyperboloidal focal surface, rather than a flat focal plane, Giacconi *et al.* (1969) determined that as the angular distance from the optical axis increased the blur circle diameter could be reduced. More recently, Nousek *et al* (1987) used this technique to optimize the *Chandra* focal plane. Much work has been done by the x-ray group at Osservatorio Astonomico di Brera in optimizing the performance and minimizing the weight of replicated parabolic-hyperbolic shells (cf. Chincarini *et al* 1998; Citterio *et al* 1998; Citterio *et al* 1999; Ghigo *et al* 1999; Conconi & Campana 2001; Conconi *et al* 2004).

In order to optimize the weight of grazing incidence mirrors, research into thin, conical foil x-ray mirrors instead of replicated shells has been made (cf. Jalota 1988; Serlemitsos 1988; Serlemitsos *et al* 1995; Petre *et al* 1999; Inneman, Pina, & Hudec 2002). An alternate approach to optimizing grazing incidence optics is the abandonment of the parabolic-hyperbolic design and using a hyperbolic-hyperbolic model (Nariai 1987; Harvey & Thompson 1999; Harvey, Thompson, & Krywonos 2000).

Werner (1977) introduced a class of grazing incidence optics that represents a revolved surface, created by the rotation of a curve in the XZ-plane about the optical axis, as a general series expansion (polynomial).

---

[1] Department of Astronomy & Astrophysics, Pennsylvania State University, University Park, PA 16802
[2] Email: roming@astro.psu.edu
[3] Department of Marketing, Pennsylvania State University, University Park, PA 16802
[4] BAE Systems Control, 600 Main St, Room S76, Johnson City, NY 13790
[5] Raytheon Company, 7700 Arlington Blvd, MS M113, Falls Church, VA 22042
[6] USDA ARS, PO Box 5367, Mississippi State, MS 39762-5367

By exploiting these polynomial surfaces, he demonstrated that relatively good resolution (FWHM < 15 arcsec) could be reached for large grazing angles. Burrows, Burg, and Giacconi (1992; hereafter BBG), using this polynomial method, demonstrated that their simulated mirror design delivered 2.5 arcsec resolution images within a field-of-view (FOV) of one degree. As-built mirrors, with optimized polynomials applied to the surface, produced 10 arcsecond resolution images at 1.5 keV over a FOV of 30 arcmin radius (Conconi *et al* 2004). BBG point out that polynomial optics are as easy to manufacture as Wolter type I mirror shells and they produce better off axis images.

Each method above provides advantages and disadvantages in optimizing grazing incident mirrors. However, because of the overall promise of the polynomial method, we have investigated the optimization of polynomial coefficients applied to a Wolter type I design in order to minimize the angular resolution of the mirror over a large grazing angle. Because the angular resolution is directly influenced by the polynomial coefficients in the general series expansion, the optimization of the polynomial coefficients can improve the angular resolution despite a wide range of grazing angles. Nevertheless, determining the optimal polynomial coefficients within a sizeable N-dimensional coefficient space can be challenging. The objective is to efficiently optimize the polynomial coefficients without probing the entire N-dimensional coefficient space (Roming *et al* 2001; hereafter Paper1).

The main technique of choice for optimizing this problem is the Metropolis-Coupled Markov-Chain Monte-Carlo (MC-MCMC) method. A description of this method can be found in Section 3.0. The Markov-Chain Monte-Carlo (MCMC) method has been used in some astronomical problems that: explore the high-dimensional parameter spaces of simulated interferometric Sunyaev-Zel'dovich effect and cluster weak lensing data (Marshall, Hobson, & Slosar 2003), fit the best parameters that model the cosmic microwave background (Chu, Kaplinghat, & Knox 2003; Verde *et al* 2003), reconstruct the emission measure distribution and the elemental abundances of the x-ray spectra of Capella (Argiroffi, Maggio, & Peres 2003), address the pile-up of the *Chandra* ACIS-S CCD (Yu *et al* 2000), and derive a baryon density and temperature profile used to construct a dark matter profile (Arabadjis & Bautz 2002). Although discussions of using MC-MCMC have been made elsewhere in other disciplines (cf. Gupta, Kilcup, & Sharpe 1988; Lill & Broughton 1992; Kim-Hung & Ferguson 1995; Andrec & Prestegard 1998; & Danese et al. 2001), this paper is the first complete discussion known to the authors of the MC-MCMC method being used in an astronomical problem.

In this paper, we present several approaches with which to contrast the optimization of polynomial coefficients against the MC-MCMC. In Section 2, we introduce the contrasting techniques and methods used for optimization. These techniques include: fractional factorial, Box-Behnken and central composite designs; and back-propagation, general regression and general method of data handling neural networks. In Section 3 we discuss the MC-MCMC method and provide an overview of the algorithm. In Section 4, we discuss our findings. In Section 5, we summarize our results and future strategies.

## 2. CONTRASTING TECHNIQUES & METHODS

An author-modified version of Interactive Ray Trace[7] was used for all ray-tracing. The design features Wolter type I mirror shells with polynomial perturbations applied to the grazing incidence surface. The modeled telescope consisted of three grazing incidence mirror shells with a 3.1 square degree FOV. For this study, only the outer shell was analyzed. For all simulations the number of source photons was chosen between 2500 and 25,000 with all photon energies set at 1.46 keV. To provide a single fitness or merit value, the ray-tracing program summed all the image spot sizes (in arcseconds) for grazing angles of 0, 1, 5, 10, 15, 20, 25, 30, 35, 40, 45, 50, 55, and 60 arcminutes. The goal was to achieve an rms image spot size of 15 arcsec or less at each of the sampled grazing angles. If the image spot size for each sampled grazing angle equals 15 arcsec then the merit value would equal the threshold of 210 arcsec. The lower the merit value is the better the performance of the mirrors. To avoid runtime errors in the program, any merit value over 1000 was automatically set to 1000.

---

[7] Interactive Ray Trace (IRT®) is a registered trademark of Parsec Technology, Inc., Boulder, CO 80301

2.1. Mirror Shell Characteristics

The baseline design was calculated using the equations from VanSpeybroeck and Chase (1972) for a coaxial and confocal paraboloid and hyperboloid. The telescope focal length is 190.5 cm. The front and back parabolic radii, as measured from the optical axis, are 196.7 mm and 190.5 mm respectively. The front and back hyperbolic radii, as measured from the optical axis, are 190.5 mm and 168.4 mm respectively. The parabolic and hyperbolic shell lengths, as measured in projection from the optical axis, are 254 mm each.

2.2. Baseline Performance without Polynomial

In order to supply a baseline to contrast future results, we first performed ray tracing on the mirror shell in which the mirror face had no polynomial perturbation utilized. It has been established that the on-axis angular resolution of Wolter type I mirror shells is excellent (see BBG). The rms spot size or resolution for the mirror shell, with no adjustments to the focal plane, either towards or away from the mirror (i.e. no defocusing - $dz$), is $3.5 \times 10^{-5}$ arcsec (This assumes an ideal surface with no scattering and no slope errors).

As the off-axis angle increases the rms spot size increases. The resolution crosses the threshold of 15 arcsec at an off-axis angle of 19 arcmin (see No Poly [dz=0] in Figure 1). By defocusing, the total performance of the mirror is improved. The optimal defocusing case (dz = 1.2 mm) is shown (No Poly [dz=1.2] in Figure 1). The resolution intersects the threshold at an off-axis angle of 25 arcmin. Although this is an improvement, the objective is to achieve a resolution of less-than 15 arcsec for all off-axis angles up to 60 arcmin.

2.3. Baseline Performance with Polynomials

Next we optimized eight polynomial coefficients, four applied to the paraboloid and hyperboloid respectively, using the downhill simplex method (DSM; see Press *et al* 1992 for a description of the DSM). Using the DSM further enhanced the performance of the mirror shell (DSM [dz=0] in Figure 1). The resolution crosses the threshold at an off-axis angle of 32 arcmin. An additional decline in the rms spot size versus off-axis angle is acquired by defocusing (Poly [dz=0.2]). Using the defocused DSM, the resolution intersects the threshold at a grazing angle of 34 arcmin. A comparison of the rms spot size found with the DSM and the no-polynomial-no-defocusing method reveals an improvement in angular resolution by almost a factor of two. Because of this improvement in angular resolution using the DSM, this method is the new baseline against which all further simulations are compared.

Despite the DSM being exceptionally robust in finding a minimum, it typically fails when searching for a global minimum in a space with many minima. The minimum found is frequently dependent on the initial input. Since the coefficient space is 8-dimensional, verifying whether the minimum is global or local is challenging. Efficiently locating a set of polynomial coefficients that provide the minimum resolution over the full FOV is still a challenge.

2.4. Statistical Design and Analysis

Rather than the brute-force hit-and-miss practice, a more economical way of optimizing a design is through the use of statistical techniques. Statistical techniques used in this study include factorial designs, of which fractional designs are a subset; and response surface designs, which include Box-Behnken and central composite designs. These methods generate a mathematical model that can be used to determine the optimal response or resolution. All statistical designs, except central composite, were created using MINITAB[8]. Central composite designs were created using an author-written program, which is based on the algorithms by Khuri and Cornell (1996; hereafter KC96). A brief discussion of each of these methods and the values used in the design is found below[9].

---

[8] MINITAB® is a registered trademark of Minitab, Inc., State College, PA 16801
[9] A complete discussion of these methods and the meaning of the design values is outside the scope of this paper. The interested reader is referred to KC96, Minitab (1997), & Ward Systems Group (1996).

## 2.4.1. $2^9$ Fractional Factorial Designs

In $2^k$ factorial designs, the input variables are calculated at two levels that are coded as –1 for the lowest value and +1 for the highest value. A design matrix of $2^k$ rows is created by performing a permutation of all possible levels of the $k$ factors, or input variables (KC96). For $k$ equal to nine, this produces a 9 x 512 matrix. Fractional factorial designs allow a fraction of the space to be sampled while preserving sufficient information to model the response. This is a substantial savings in computational time.

Test runs using a ¼ fractional factorial design were explored. The highest and lowest values for the polynomial coefficients used in the design can be found in Roming *et al.* (2000; hereafter Paper2). The design was composed of nine factors[10], 128 runs, no center points, no blocking, one replicate, and a resolution of VI (see KC96 for a description of the design parameters).

## 2.4.2. Box-Behnken designs

The Box-Behnken approach is a class of three-level partial factorial designs for ascertaining parameters in a second-order model. Merging two-level factorial designs and balanced incomplete block designs in a systematic way form the Box-Behnken designs (KC96). The design was composed of 62 runs, no blocking, six center points, and seven factors. Due to limitations in MINITAB seven factors were used instead of nine.

Some test runs were optimized using a genetic algorithm. All designs optimized with a genetic algorithm utilize GeneHunter[11] (see Ward Systems Group 1995). The genetic algorithm varied the seven parameters and evolved until a minimum was found. Each algorithm consisted of a population size of 2000, chromosome length of 32 bits, and was run until no new minimum was encountered for 50 generations.

## 2.4.3. Central composite designs

An alternative to $3^k$ factorial designs is the central composite design. The design is composed of three parts: a complete or fraction of a $2^k$ factorial design; one or more center points; and two axial points on the axis of each design variable at a distance of $\alpha$ from the center point (KC96).

The design comprised nine factors, 36 center points, an $\alpha$-value of 4.757, 566 runs, and no blocking. As in the Box-Behnken case, some test runs were optimized using a genetic algorithm with the same parameters as above.

## 2.5. Artificial Neural Networks (ANN)

An alternate optimization tool to statistical methods is artificial neural networks (ANNs). For many applications, an ANN successfully replicates complex relationships that exist between inputs and outputs without the need for an equation or model *a priori*. An artificial neural network is a data processing system consisting of a large number of simple, highly interconnected processing elements (artificial neurons) in an architecture inspired by the structure of the cerebral cortex of the brain (Tsoukalas and Uhrig, 1997). In simple terms, the artificial neuron is analogous to the biological neuron in that it has inputs (dendrites), a processing unit (soma) and output(s) (axons). The output of one neuron can connect with the input of another. Thus, an ANN appraises every interaction connecting elements and creates relationships between variables (Roush & Cravener 1997). ANNs employed in this study are back-propagation neural networks (BPNNs), general regression neural networks (GRNNs), and general handling of data method neural networks (GMDH). For this study, all ANNs were created using NeuroShell 2[12]. An analysis of the individual results of some test runs, representing different regions of the N-dimensional coefficient space, can be found in Paper2.

---

[10] The number of factors was increased from eight to nine with the intent of maximizing the performance of the algorithms. If a factor was unimportant the algorithm should find it to be zero. As will be shown later in the paper, this increase from eight to nine did not improve the performance of the 'contrasting' methods.
[11] GeneHunter® is a registered trademark of Ward Systems Group, Inc., Frederick, MD 21702
[12] NeuroShell® 2 is a registered trademark of Ward Systems Group, Inc., Frederick, MD 21702

2.5.1.  Back-propagation Neural Network (BPNN)

The most popular and extensively used ANN is the BPNN (Chester 1993). The BPNN is a supervised, multi-layered, feed-forward network that utilizes the Back Propagation Rule. This rule is founded upon the Delta Rule (Chester 1993; Tsoukalas & Uhrig 1997) for networks with hidden layers. Training includes moving patterns forward from the beginning to the end of the network layers, then propagating the errors backward, and then updating the weights. This is done to diminish the errors. The mean square error is the value being minimized (Lawrence 1993).

The network learns by a process involving the modification of connection weights between neurons. The artificial neuron receives inputs and corresponding output values. When information is loaded into an ANN, it must be scaled from the current numeric range into a range that the ANN deals with efficiently. Linear scales of [0, 1] and [-1, 1] are most commonly used. After variables are imported into the ANN and scaled, a calibration set is extracted for use during ANN training. The percentage of the database extracted and the method of pattern extraction may be altered. A pattern is a single row of the database or a single observation. A rotation method of extraction selects patterns in the order they appear in the database. A random method randomly chooses the calibration patterns.

When a network with back-propagation architecture is presented with a training set, the neuron transforms sums of inputs into weights that are transferred to other neurons. The difference between the predicted and the actual training output is computed. The error is propagated backwards through the hidden layer to the input layer. The connection weights between neurons and layers are adjusted until the output error is minimized. The amount of adjustment is determined by the learning rate multiplied by the error. The momentum weight is associated with the learning rate. The momentum weight not only includes the change dictated by the learning rate, but also includes a portion of the last weight change. With the TurboProp[13] method (Ward Systems Group 1996), weights are only updated after all patterns in the database have been evaluated. All of the training set data are presented until the network is able to duplicate the training set with success. To avoid over training the neural network, the ANN is evaluated against a test set during the training process. After training, the ANN can then be used to predict outputs when provided inputs upon which the network has not been trained.

For our BPNN, the number of neurons in the input layer was nine, corresponding to the number of polynomial coefficients. The hidden layer was composed of three slabs. There were fifteen neurons in each of the hidden slabs. The number of neurons in the output layer was one, corresponding to the calculated resolution. The scale function for the input layer was a linear function. This allowed for extrapolation outside the sampled region. A Gaussian, hyperbolic tangent, and Gaussian complement scale function was used for the hidden slabs. The output layer scale function was linear (see Figure 2). The learning rate and momentum were each 0.1; a value of 0.3 was applied for the initial weight. The training criterion used can be found in Paper2.

2.5.2.  General Regression Neural Network (GRNN)

GRNN are a feed-forward ANN established on the theory of non-linear regression. GRNN training of the pattern units is supervised, i.e. the network is shown how to make predictions by the user supplying a large number of accurate predictions (Masters 1995). This training exploits a unique algorithm that makes characterizing the number of pattern units *a priori* unnecessary (Tsoukalas & Uhrig 1997). These networks train swiftly on sparse data sets and are valuable for continuous function approximation. They are capable of taking multidimensional inputs and replicating multidimensional surfaces (Roush & Cravener 1997). The GRNN is a three-layer network wherein one hidden neuron must be present for each training pattern (Masters 1995)

---

[13] TurboProp[TM] is a trademark of Ward Systems Group, Inc.

The GRNN functions by measuring how far an input prediction is from an N-dimensional space training calibration set output, where N is the number of inputs in the problem. When a new pattern is presented to the network, it is compared in N-dimensional space to all of the patterns in the training set to determine how far in distance it is from those patterns. Two methods of calculating this distance metric are used: the Euclidean and the City Block methods. The Euclidean method is based on the formula for the hypotenuse of a right triangle. That is, the Euclidean distance is the square root of the sum of the differences squared in all dimensions between the pattern and the weight vector for a neuron. The City Block distance is the sum of the absolute values of the differences in all dimensions between the pattern and the weight vector of a neuron. It is the sum of the legs of a right triangle (Varmuza 1980). The success of GRNN is dependent on a smoothing factor. For quantitative work the smoothing factor defines the shape of the path of data responses. Small smoothing factors closely connect each data point. As the smoothing factor increases, the path through all data points becomes more generalized; until at the extreme, it no longer provides useful information about the shape of the responses. A genetic algorithm (i.e. genetic adaptive calibration) can be used to find an appropriate smoothing factor. Masters (1996) gives a detailed description of the form and function of a GRNN.

For our GRNN, the number of neurons in the input layer was nine, mapping to the number of polynomial coefficients. The hidden layer was composed of 1596-1600 neurons. The number of neurons in the output layer was one, corresponding to the predicted resolution (see Figure 3). A description of the training criteria for most runs is the same as Run #1 in Paper 1. The training criterion for the remaining runs is as described in Paper 1. No improvement was encountered when smoothing factors were varied from 0.1 to 0.5. Clipping of the maximum allowed spot size for some of the test runs was also tried to determine if the ANN would model the data better. Genetic algorithms were used to minimize the rms spot size.

2.5.3. General Method of Data Handling (GMDH)

GMDH is sometimes referred to as the "polynomial net." It is developed by building successive layers with complex links that are the individual terms of a polynomial. These terms are created by linear and nonlinear regression (Farlow 1984). The initial layer is the input layer. Computing regressions of the input variables and then selecting the best variables create the first layer. The second layer is created by computing regressions of the values from the first layer along with the input variables. Only the best variables remain in the polynomial. The process continues until there is no longer improvement in the net. The resulting ANN is a complex polynomial that describes the nature of the relationship of the inputs and outputs. One of advantage of the GMDH is that "prejudiced" mathematical models are not requisite, simply the data (Farlow 1984).

For our GMDH ANN, the number of neurons in the input layer was nine, corresponding to the number of polynomial coefficients. The constructed number of hidden layers ranged from 21 to 89. The number of neurons in the output layer was one, representing the predicted resolution. The scale function in both the input and output layers was a linear function that allowed extrapolation outside the modeled area (see Figure 4). The training criteria can be found in Paper 1. In some cases, the training pattern was increased to 2000.

### 3. METROPOLIS-COUPLED RANDOM-WALK HASTINGS-METROPOLIS MARKOV-CHAIN MONTE-CARLO METHOD

Recent work in statistics has led to the Markov-Chain Monte-Carlo (MCMC) methodology for performing numerical integrals with respect to high dimensional probability distributions (Gilks, Richardson, & Spiegelhalter 1996). At the core of the MCMC methodology is an algorithm known as the Hastings Metropolis (HM) algorithm (Hastings 1970; Metropolis, *et al* 1953). MCMC algorithms use the HM algorithm to create a Markov chain on the parameter space of a particular probability or target function. This Markov chain converges in distribution to the probability function of interest. Once the Markov chain has converged, draws from the chain can be used to construct an estimate of the expectation of arbitrary functions with respect to this probability distribution.

The Markov chains which result from an MCMC algorithm have several appealing properties which should make them desirable tools for searching a parameter space for a global maximum/minimum of an arbitrary function. The Markov chain that results from a MCMC algorithm can be viewed as a type of "random walk" with drift over a parameter space. The drift is in the direction of larger/smaller values of the desired function. If constructed properly, the resulting Markov chain will be irreducible. This means given enough time, it has a positive probability of visiting every part of the parameter space. Combining the fact that the chain tends to drift in a positive/negative direction with the fact that it is irreducible means that the chain will spend most of its time in regions with high/small values, but not get stuck in local maximums/minimums. In order to ensure that the chain that was used to optimize the function discussed in this paper did not get caught in a local maximum/minimum, we used a version of the MCMC algorithm known as the Metropolis-coupled MCMC (MC-MCMC) algorithm (Geyer 1991).

The MC-MCMC algorithm runs $N$, Random Walk, Markov Chains in parallel and randomly switch values between the different chains. To be precise, let $\pi_n(\theta^n) = (\pi_0(\theta^n))^{1/(1+\lambda(n-1))}$, with $\lambda > 0$, where $\pi_0(\theta)$ is the target density[14] of interest. The goal is to generate a Markov chain, $\theta_t^n$, for each stationary density $\pi_n(\theta^n)$, but to borrow from all of the chains in an appropriate manner. Each Markov chain is updated in two ways: first its current parameter values are randomly perturbed $\theta^* \sim Q(\theta^* | \theta_t^n)$ and the proposed value is accepted with probability

$$\alpha(\theta_t^n, \theta^*) = \min\left\{ \frac{Q(\theta_t^n | \theta^*)\pi_n(\theta^*)}{Q(\theta^* | \theta_t^n)\pi_n(\theta_t^n)}, 1 \right\},$$

(or with probability $\alpha$, $\theta_{t+1}^n = \theta^*$, otherwise $\theta_{t+1}^n = \theta_t^n$) which is a standard Random Walk, Hasting Metropolis updating move; and second two chains are randomly chosen and the parameters from both chains are switched with probability

$$\alpha(switch(i,j)) = \min\left\{ \frac{\pi_i(\theta_t^j)\pi_j(\theta_t^i)}{\pi_i(\theta_t^i)\pi_j(\theta_t^j)}, 1 \right\},$$

(or with probability $\alpha$, $\theta_{t+1}^j = \theta_t^i$ and $\theta_{t+1}^i = \theta_t^j$, otherwise everything remains unchanged). For a discussion of the exact implementation of this algorithm that was used, see Paper1. The first type of moves, standard Random Walk moves, allows each chain to traverse the parameter space independently. One of the key elements of this strategy is that as $n$ increases the accompanying stationary density becomes more and more 'flat' (in contrast to the target density $\pi_0(\theta)$); which allows the accompanying chains to move more and more rapidly around the parameter space. The switching step then allows the target chain to try to move to parts of the parameter space proposed by the other stationary densities, which are relatively far from its current location, resulting in rapid exploration of complex, high dimensional functions.

---

[14] Note that $\pi_0(\theta)$, must be a strictly positive, integrable function and in a typical statistical application the target density would be a posterior probably density. In cases where the function is not integrable, the MCMC algorithm is not guarantied to converge in distribution as the resulting Markov chain is no longer irreducible, but, as demonstrated by this application, the MCMC algorithm may still prove to be a useful tool for efficiently exploring and identifying maximum/minimum of complex functions.

4. RESULTS

The total merit value for each method is listed in Table 1. If a method was able to produce a total merit value of 15 arcsec over the entire grazing angle space (0-60 arcmin) then the summed merit would be 210, i.e. the threshold. The polynomials from the run that produced the best results for each method were applied to the design and the image spot size at each off-axis angle was obtained (see Figure 5). The merit values for the Central Composite and GRNN designs were extremely high; therefore they are not plotted in Figure 5. To provide greater clarity, the No poly designs and the DSM ($dz = 0$) designs are also not plotted in Figure 5. Both the Central Composite and the GRNN approach did a very poor job of modeling the polynomial space. The BPNN and Fractional Factorial techniques produced approximately the same results as the No poly ($dz = 0$) case. The Box-Behnken and GMDH procedures performed similar to the defocused No poly method.

From Figure 5 and Table 1, it is evident that the two best approaches are the DSM and the MC-MCMC methods. Both runs of the MC-MCMC method produced merit values almost half that of the defocused DSM method. Because of the excellent results from the MC-MCMC method, the two best results are plotted in the Figure. For the best solution of the MC-MCMC method, the spot size is 1.75, 8.44, 10.28, 16.98, and 57.20 arcseconds at 0, 15, 30, 45, and 60 arcminute off-axis angles respectively. As expected, the larger the off-axis angle the more pronounced the aberration. The polynomial applied to the mirror is as follows: $p_0 = 9.29 \times 10^{-1}$, $p_1 = -2.45 \times 10^{-2}$, $p_2 = -7.54 \times 10^{-4}$, $p_3 = -7.19 \times 10^{-6}$, $p_4 = 0.00 \times 10^{0}$, $h_0 = -1.33 \times 10^{+1}$, $h_1 = 894 \times 10^{-2}$, $h_2 = 8.72 \times 10^{-4}$, $h_3 = -3.06 \times 10^{-6}$, & $h_4 = -1.85 \times 10^{-9}$. The values $p_0$ to $p_4$ and $h_0$ to $h_4$ represent the zeroth to fourth order polynomial coefficients for the paraboloid and the hyperboloid respectively.

5. SUMMARY

We have modeled grazing incidence optics with a polynomial perturbation applied to the mirror surface. The objective was to find a technique that efficiently optimizes the polynomial coefficients that affect the minimization of the angular resolution. The techniques explored included: fractional factorial designs; response surface, specifically Box-Behnken and central composite designs; artificial neural networks, particularly back-propagation, general regression, and general method of data handling neural networks; and a Metropolis-Coupled Markov-Chain Monte-Carlo algorithm. All designs were tested over a large region of multidimensional coefficient space. The results were contrasted against a Wolter type I design and a Wolter type I design with polynomial perturbations that were optimized using the Downhill Simplex Method.

Most of the methods were found to be worse than or comparable to the DSM. Most of the methods we investigated suffer from the same difficulty that the DSM suffers from: they often fail when searching for a global minimum in a space with many minima. The minima found by these methods depend on the initial inputs. The utilization of the MC-MCMC method means that chains will tend to not get caught in local minimums. The MC-MCMC method was found to be very good at minimizing the angular resolution over a wide range of grazing angles.

At 15 arcsecond resolution these mirrors can resolve $10^{3.9}$ sources/degree$^2$. This source density occurs at logS = -16.6 erg/cm$^2$/s and logS = -15.5 erg/cm$^2$/s for the soft (0.5-2 keV; see Moretti *et al* 2002) and hard (2-8 keV; see Cowie *et al* 2002) x-ray bands, respectively. In the soft x-ray band, this is 48 times and 12 times deeper than the ROSAT Ultra Deep Survey (Lehmann *et al* 2001) and the deep *XMM*-Newton survey of the Lockman Hole (Hasinger *et al* 2001) respectively. In the hard x-ray band, this is four times deeper than the deep *XMM*-Newton survey of the Lockman Hole. Thus, a single maximally deep unconfused pointing of a telescope with these mirrors will have $10^{4.4}$ sources, and 2000 pointings will generate $10^{7.7}$ sources and would cover 15% of the celestial sphere.

ACKNOWLEDGMENTS

We would like to express our gratitude to Eric Feigelson for his encouragement and excellent comments that have improved this work. We would also like to thank Lisa Rabban for her assistance in the data analysis. We appreciate the support from Scott Koch and Pat Broos in coding the necessary algorithms. This work has been funded by NASA grant NAG5-5093.REFERENCES

Andrec, M., & Prestegard, J. H. 1998, Journal of Magnetic Resonance, 130, 217
Arabadjis, J. S., & Bautz, M. W. 2002, BAAS, 34, 1094
Argiroffi, C., Maggio, A., & Peres, G. 2003, A&A, 404, 1033
Burrows, C. J., Burg, R., & Giacconi, R. 1992, ApJ, 392, 760 [BBG]
Chincarini, G., Citterio, O., Conconi, P., Ghigo, M., & Mazzoleni, F. 1998, Astron. Nachr., 319, 125
Chester, M. 1993, Neural Networks: A Tutorial (PTR Prentice Hall, Englewood Cliffs)
Chu, M., Kaplinghat, M., & Knox, L. 2003, ApJ, 596, 725
Citterio, O., Campana, S., Conconi, P., Ghigo, M., Mazzoleni, F., Braeuninger, H. W., Burkert, W., & Oppitz, A. 1998, SPIE, 3444, 393
Citterio, O., Campana, S., Conconi, P., Ghigo, M., Mazzoleni, F., Braeuninger, H. W., Burkert, W., & Oppitz, A. 1999, SPIE, 3766, 198
Conconi, P., & Campana, S. 2001, A&A, 372, 1088
Conconi, P., Pareschi, G., Campana, S., Chincarini, G., & Tagliaferri, G. 2004, SPIE, 5168, 334
Cowie, L. L., Garmire, G. P., Bautz, M. W., Barger, A. J., Brandt, W. N., & Hornschemeier, A. E. 2002, ApJ, 566, L5
Danese, G., De Lotto, I., De Marchi, A., Leporati, F., Bellini, T., Buscaglia, M., & Mantegazza, F. 2001, Computer Physics Communications, 134, 47
Djuric, P. M. 1998, SPIE, 3459, 262
Farlow, S. J. 1984, Self-Organizing Methods in Modeling (Marcel Dekker, Inc., New York)
Fox, C., Palm, M., & Nicholls, G. K. 1999, SPIE, 3816, 23
Geyer, C. J. 1991, in Computing Science and Statistics: Proceedings of the 23[rd] Symposium of the Interface, ed. E. M. Keramidas (Fairfax Station: Interface Foundation), 156
Ghigo, M., Citterio, O., Mazzoleni, F., Kolodziejczak, J. J., O'dell, S. L., Austin, R. A., & Zirnstein, G. 1999, SPIE, 3766, 207
Giacconi, R., Reidy, W. P., Vaiana, G. S., VanSpeybroek, L. P., & Zehnpfennig, T. F. 1969, Space Science Reviews, 9, 3
Gilks, W. R., Richardson, S., & Spiegelhalter, D. J. 1996, Markov Chain Monte Carlo in Practice (Chapman & Hall, London)
Gupta, R., Kilcup, G. W., & Sharpe, S. R. 1988, PhRvD, 38, 1278
Harvey, J. E., & Thompson, P. L. 1999, SPIE, 3779, 371
Harvey, J. E., Thompson, P. L., & Krywonos, A 2000, SPIE, 4012, 328
Hasinger, G., *et al* 2001, A&A, 365, L45
Hastings, W. K. 1970, Biometrika, 57, 97
Inneman, A. V., Pina, L., & Hudec, R. 2002, SPIE, 4496, 73
Jalota, L. 1988, Ph.D. Thesis, Leicester University (England)
Khuri, I., & Cornell, J. A. 1996, Response Surfaces (Marcel Dekker, Inc., New York) [KC96]
Kim-Hung, C., & Ferguson, D. M. 1995, Computer Physics Communications, 91, 283
Lawrence, J. 1993, Introduction to Neural Networks (California Scientific Software Press, Nevada City)
Lehmann, I., *et al* 2001, A&A, 371, 833
Lill, J. V., & Broughton, J. Q. 1992, PhRvB, 46, 12068
Marshall, P. J., Hobson, M. P., Slosar, A. 2003, MNRAS, 346, 489
Masters, T. 1995, Advanced Algorithms for Neural Networks (John Wiley & Sons, Inc., New York)
Metropolis, N., Rosenbluth, A. W., Rosenbluth, M. N., Teller, A. H., & Teller, E. 1953, J. Chem. Phys., 21, 1087
Minitab Inc. 1997, MINITAB User's Guide 2: Data Analysis and Quality Tools (Minitab Inc., State College, PA)
Moretti, A., Lazzati, D., Campana, S. & Tagliaferri, G. 2002, ApJ, 570, 502
Nariai, K. 1987, Applied Optics, 26, 4428
Nicholls, G. K., & Fox, C. 1998, SPIE, 3459, 116
Nousek, J. A., Garmire, G. P., Ricker, G. R., Bautz, M. W., Levine, A. M., & Collins, S. A. 1987, SPIE, 818, 296

| Methodology | Summed Merit |
|---|---|
| No poly ($dz = 0$) | 678 |
| No poly ($dz = 1.2$) | 608 |
| DSM ($dz = 0$) | 428 |
| DSM ($dz = 0.2$) | 414 |
| Fractional Factorial | 738 |
| Box-Behnken | 628 |
| Central Composite | <20k |
| BPNN | 681 |
| GRNN | 20,000 |
| GMDH | 583 |
| MC-RWHM-MCMC (1) | 282 |
| MC-RWHM-MCMC (2) | 217 |
| Threshold | 210 |

Table 1. Total merit value for each methodology.

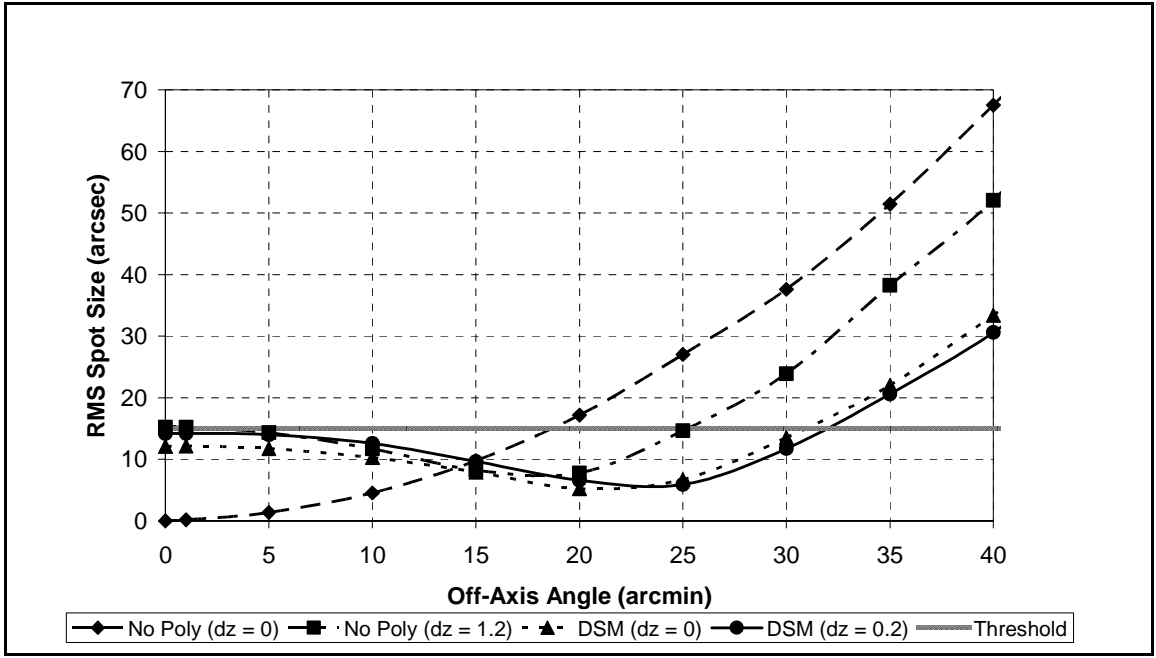

Figure 1. A comparison plot of the rms spot sizes produced by Wolter I designs with no polynomial applied and Wolter I designs with a polynomial applied in which the polynomial coefficients were optimized using the Downhill Simplex Method (DSM).

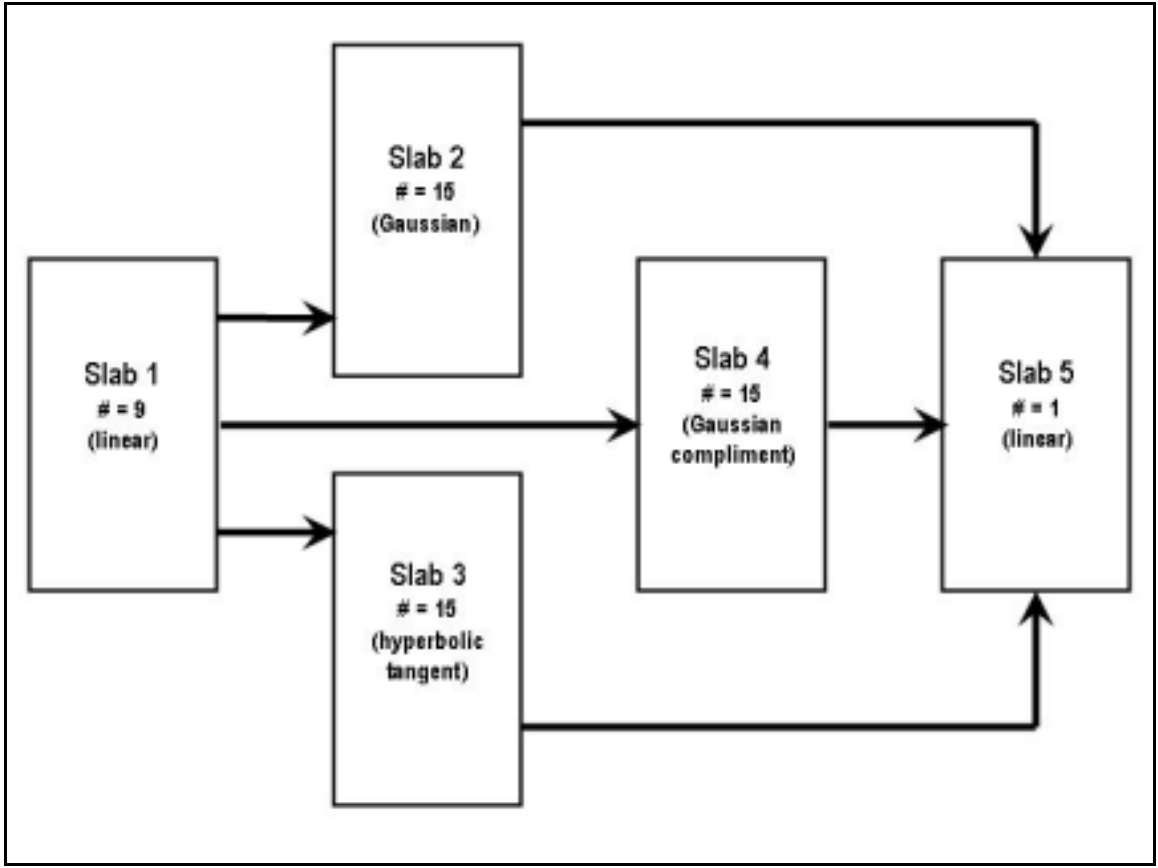

Figure 2. Back-Propagation Neural Network (BPNN) architecture used by the authors. This neural network was developed by the Ward Systems Group (1996). The number (#) represents the number of neurons in the given slabs. The values in parenthesis are the scale functions.

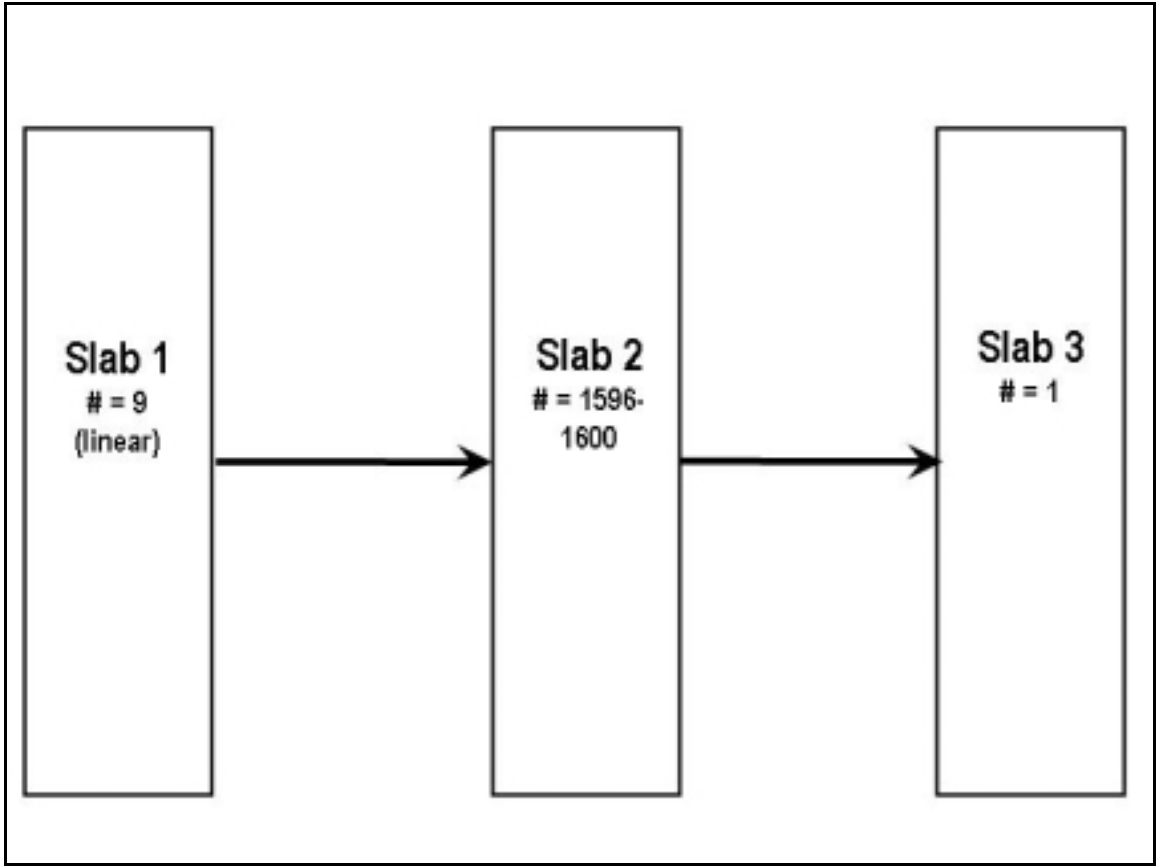

Figure 3. General Regression Neural Network (GRNN) architecture used by the authors. The number (#) represents the number of neurons in the given slabs. The value in parenthesis is the scale function for the input slab.

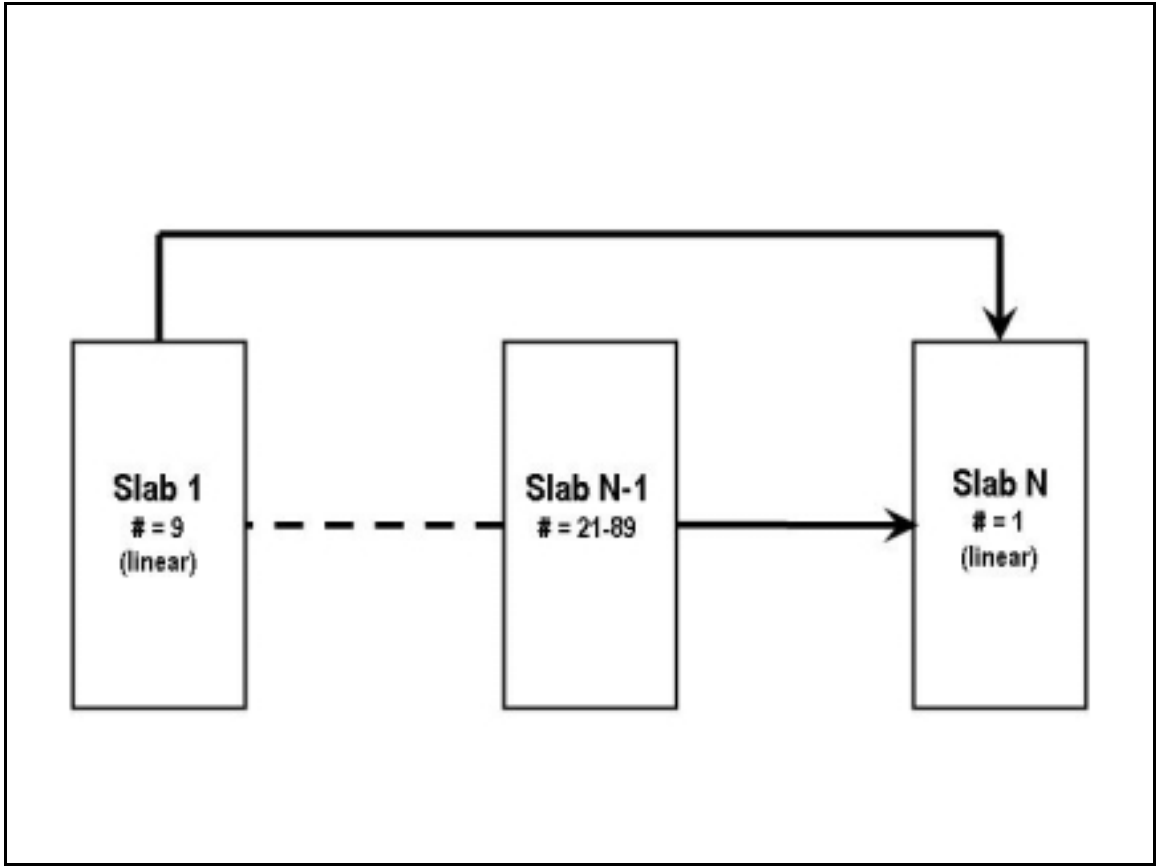

Figure 4. Group Method of Data Handling (GMDH) Neural Network architecture used by the authors. The number (#) represents the number of neurons in the given slabs. The values in parenthesis are the scale functions for the input and output slabs.

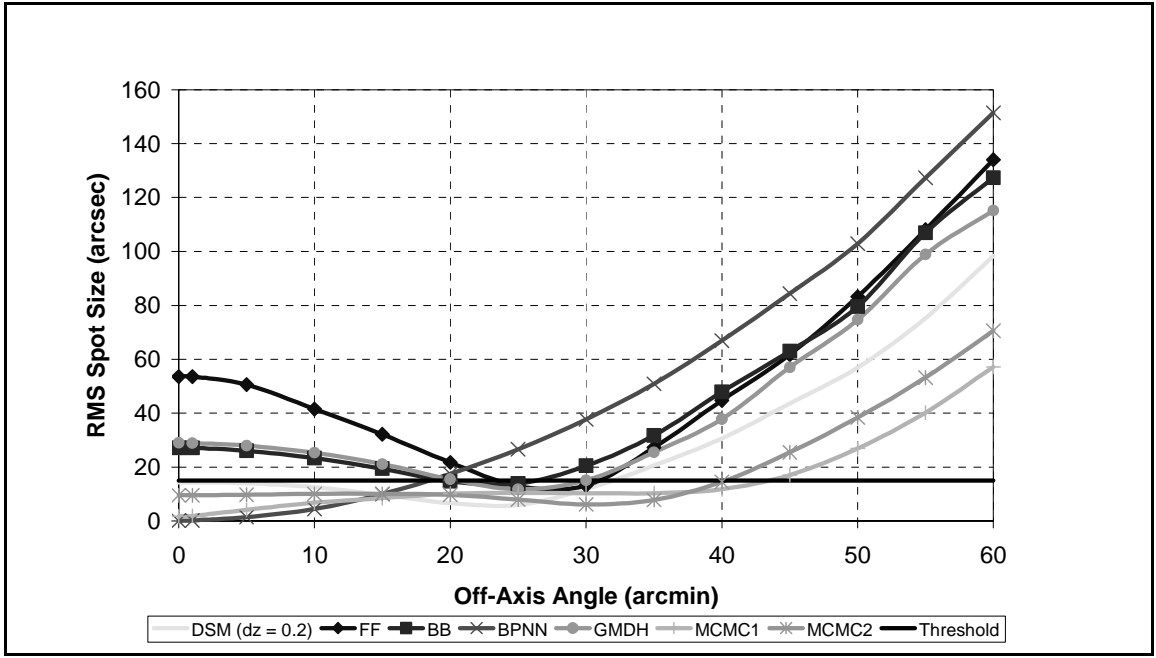

Figure 5. Plot of the merit value versus off-axis angle for each optimization method. For clarity, the No poly, DSM ($dz = 0$), Central Composite, & GRNN designs are not plotted.